\documentclass[prl,showpacs,twocolumn,amsmath,amssymb,aps]{revtex4-1}

\usepackage{amssymb}
\usepackage{graphicx}

\begin{document}

\title{One-Dimensional Photon Transport Through a Two-Terminal Scattering Cluster: Tight-Binding Formalism}

\author{Yu Jiang}
\affiliation{Departamento de F\'{\i}sica, Universidad Aut\'onoma
Metropolitana-Iztapalapa, A. P. 55-534, 09340 Mexico City, Mexico}
\author{M. Lozada-Cassou}
\affiliation{Instituto de Energías Renovables, Universidad Nacional Autónoma de México (UNAM), 62580 Temixco, Morelos, México}
\
\begin{abstract}
Employing tight-binding approximation we derive a transfer matrix formalism for one-dimensional single photon transport through a composite scattering 
center, which consists of parallel connected resonator optical waveguides. By solving the single-mode eigenvectors of the Hamiltonian, we investigate
the quantum interference effects of parallel couplings on the photon transport through this parallel waveguide structure. We find a perfect reflection
regime determined by the number of coupled resonator waveguides. Numerical analysis reveals that by changing atom transition frequency, the window of
perfect reflection may shift to cover almost all incoming photon energy, indicating the effective control of single photon scattering by photon-atom 
interaction.
\end{abstract}


\maketitle

\section{I. Introduction}

Photon transport in one dimensional coupled waveguide is an important model system for exploring quantum information processing and manipulation mechanism.
The coupling of atoms or quantum emitters to the optical waveguide offers feasible control schemes to achieve quantum switching, routing, photon storages
and other quantum information operations\cite{Srini, Raimond, Pelton, Bermel, Menon, Astaf}. Currently there are several important theoretical approaches
being used to study the photon transport in one dimension, including the real space Bethe anzatz approach\cite{Shen05, Shen07}, Input-out formalism 
\cite{Gardiner, Fan10}, and Lippman-Schwinger scattering-theory approach \cite{Fan99, Xu, Tsoi, Roy}. Recently a tight-binding formalism has been employed
to show that the scattering of a single photon inside a one-dimensional resonator waveguide can be ocntrolled by the coupled two-level quantum 
system\cite{Zhou08}. It is demonstrated that a finite-bandwidth spectrum of perfect reflection appears as the detuning between the photon and atomic 
frequency varies. It is obvious that when a single photon transport properties can be influenced by the number of atoms that interact with the waveguide 
as a result of multiple scattering between the propagating photon and the quantum emitters\cite{Yanik04, Yanik05, Zhou08a, Chang11, Zhou13, Cheng12, 
Liao16, Qin16, MZI}. In addition to the effects of the collective interaction with the multiple emitters, the coupling modes between atom and the 
one-dimensional waveguide also affect the photon transport and result in quite involved, nontrivial dispersion relations that can lead to strong 
reduction of the group velocity of photons\cite{Yanik04, Yanik05, Calajo}, as a consequence of a finite bandwidth. Since the coupling configuration 
between atoms and waveguide play an important role in determining the transport properties of photons, in this paper we will focus on the photon 
transfer through a black-box like scattering cluster, by employing the tight-binding theory\cite{Kowal}. 

This paper is organized as follows. In Sec. II we derive a general theory for photon transport through a two-terminal interaction box, where multiple 
atoms are coupled in series as well as in parallel. We then discuss the simple case of coupled cavity array based on the Jaynes-Cummings model for 
single cavities, in Sec. III. Finally we give a summary in Sec. VI.

\section{II. General Theory}
The model system discussed in this work consists of two leads sndwiched by a tight-binding network in the parallel configurations, as illustrated 
in Fig.1. 
\begin{figure}
\includegraphics[width=6.9cm]{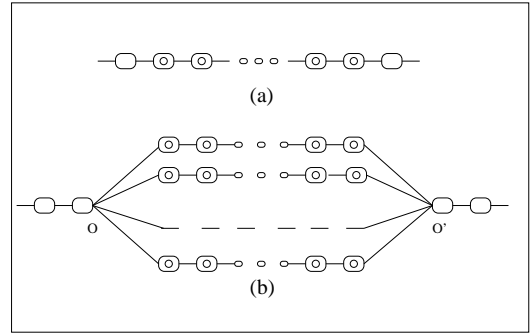}
\caption{The schematics of the single coupled resonator waveguide (a), and the two-port parallel connected multiple coupled resonator waveguides (b).}
\end{figure}

In this section we derive the transmission and reflection amplitudes in the Tight-Binding (TB) model for parallel connected transmision lines. 
The typical TB equations, resulting from the Schrodinger equation $H\Psi=E\Psi$ can be written as
\begin{equation}
\label{eq:TB}
(\epsilon_j-E)\psi(j)=\sum_nJ_{j,n}\psi(j+n)
\end{equation}
For purposes of illustration of our method, we consider here only the simplest case where the transfer integral between the nearest neightbors 
$J_{j,n}=1$ and the lattice constant $a=1$. We also assume that the scattering is elastic and the traveling wave functions may take the following 
form
\begin{equation}
\label{eq:Traveling}
\psi(j)=Ae^{ikj}+Be^{-ikj},
\end{equation} 
with the dispersion relation $\epsilon_j-E_k=2\cos(k)$.
Suppose that there are $n_j$ sites at the $j$-th channel, and the total number of the TB sites in the central scattering network is $N=\sum^{m}_{j=1}n_j$. 
Here $m$ represents the number of the parallel channels that converge into the left and right terminals.

The TB equations of the parallel configuration can be written as 
\begin{equation}
\begin{split}
	(\epsilon_j-E_k)\psi(-1) = & \psi(-2)+\sum^{n_i}_i\psi_i(0), \\
	(\epsilon_j-E_k)\psi(N) = & \psi(N+1)+\sum^{n_i}_i\psi_i(n_i-1). 
\end{split}
\end{equation}
By taking the standard expressions for the left incoming and the right outgoing waves,
\begin{equation}
\begin{split}
	\psi_L(j) = & e^{ikj}+re^{-ikj}, \quad j<0, \\
	\psi_R(j) = & te^{ikj}, \quad j>N, 
\end{split}
\end{equation}
we find
\begin{equation}
\begin{split}
	1+r = & \sum_{i=1}^{m}\psi_i(0), \\
	t_N = & \sum_{i=1}^{m}\psi_i(n_i-1), 
\end{split}
\end{equation}
where $t_N=te^{ik(N-1)}$, and $\psi_i(j)$ is the wave function at the site $j$ of the $i$-th channel.

Now let us derive a relation between the wave functions $\psi_i(0)$ and $\psi_i(n_i-1)$. To this end, we resort to the transfer matrix expression of 
the TB equations,
\begin{equation}
\begin{bmatrix}
	\psi_i(j+1)\\
	\psi_i(j)\\
\end{bmatrix}
=\begin{pmatrix}
	 \alpha_{i,j} & -1  \\
	 1 & 0 \\
\end{pmatrix}
\begin{bmatrix}
    \psi_i(j)\\
	\psi_i(j-1)\\
\end{bmatrix},
\end{equation}
with $\alpha_{i,j}=\epsilon_{i,j}-E_k$. Here $\epsilon_{i,j}$ stands for the energy of the $j$-th site of the $i$-th waveguide channel. It is 
straightforward that
\begin{equation}
\begin{bmatrix}
	\psi_i(n_i)\\
	\psi_i(n_i-1)\\
\end{bmatrix}
=M_i
\begin{bmatrix}
    \psi_i(0)\\
	\psi_i(-1)\\
\end{bmatrix}.
\end{equation}
Taking into account of the continuity condition of the wave functions at any TB sites, we have that $\psi_i(-1)=1+r$ and $\psi_i(n_i)=\psi(N)=t_N$. 
Introducing that
\begin{equation}
M_i=
\begin{pmatrix}
	 \alpha_{i,n_i-1} & -1  \\
	 1 & 0 \\
\end{pmatrix}
...
\begin{pmatrix}
	 \alpha_{i,0} & -1  \\
	 1 & 0 \\ 
\end{pmatrix}\nonumber
\end{equation}
\begin{equation}
=
\begin{pmatrix}
	 a_{i,n_i-1} & b_{i,n_i-1}  \\
	 c_{i,n_i-1} & d_{i,n_i-1} \\
\end{pmatrix}
\end{equation}
where the matrix elements can be obtained from the following recursion relations
\begin{equation}
\begin{split}
	a_{i,n}= & \alpha_{i,n-1}a_{i,n-1}-c_{i,n-1}, \\
	b_{i,n}= & \alpha_{i,n-1}b_{i,n-1}-d_{i,n-1}, \\
	c_{i,n}= & a_{i,n-1}, \\
	d_{i,n}= & b_{i,n-1},
\end{split}
\end{equation}
with $a_{i,0}=\alpha_{i,0}$, $b_{i,0}=-1$, $c_{i,0}=1$ and $d_{i,0}=0$. By inserting (7)-(9) into (5) 
we get the transmission and reflection amplitudes for the scattering cluster formed by parallel coupled oupled cavity waveguides,
\begin{equation}
\begin{split}
	r = & \frac{(P'e^{ik}-1)(1-Qe^{-ik})+PQ'}{(Qe^{ik}-1)(P'e^{ik}-1)-PQ'e^{2ik}}, \\
	t_N = & \frac{-2iQ'\sin(k)}{(Qe^{ik}-1)(P'e^{ik}-1)-PQ'e^{2ik}}, 
\end{split}
\end{equation}
with
\begin{equation}
P=\sum_{i=1}^{m}\frac{1}{a_{i,n_i-1}}, \quad Q=-\sum_{i=1}^{m}\frac{b_{i,n_i-1}}{a_{i,n_i-1}},
\end{equation}
\begin{equation}
P'=\sum_{i=1}^{m}\frac{c_{i,n_i-1}}{a_{i,n_i-1}}, \quad Q'=\sum_{i=1}^{m}(d_{i,n_i-1}-\frac{b_{i,n_i-1}c_{i,n_i-1}}{a_{i,n_i-1}}).
\end{equation}

This is the main result of this work. It is worth to point out that the physical relavance of the site energy $\epsilon_i(j)$ may be quite general, 
including the PT-symmetric potentials. It is remarkable that we can evaluate the photon transport properties without assuming the detailed wave 
functions in the scattering zone, provided that the tight-binding approximation is valid, in sharp contrast with the current approaches used in 
the literature.

\section{III. Numerical analysis}

To illustrate our method, in the following, we consider the problem of the single photon transport in an one-dimensional coupled resonator 
waveguides. To be more specific, we assume that photon propagates along the left waveguide and enters the multiple parallel coupled waveguides 
through the splitter point $O$. And then it is transfered to the left waveguide via the converter $O'$. As part of the scattering cluster those 
one-dimensional waveguide are coupled with two-level atoms each. Let us denote by $a^{\dagger}$ the single mode in the $j$-th cavity, with 
frequency $\omega$. The Hamiltonian of the CRW is given by
\begin{equation}
H_{cv} = \omega\sum_ja^{\dagger}_ja_j-\sum_{j,j'}(J_{j,j'}a^{\dagger}_ja_{j'}+H.c.)
\end{equation}
where $J_{j,j'}$ is the inter-cavity photon hopping constant among the connecting cavities. For uniform hopping constant $ J_{j,j'}=J$, the 
Hamiltonian (1) can be readily diagonalized to yield the dispersion relation $E_k=\omega-2J\cos(k)$. Here we assume that $\hbar=1$ and the 
lattice constant (inter-cavity distance) $a=1$.

We further assume that the Hamiltonian of the atom is given by $H_A=\sum\omega_j|e_j><e_j|$ and the interaction of the single photon with a 
two-level atom inside the $j$-th cavity, is then described by the Jaynes-Cummings Hamiltonian
\begin{equation}
H_j = g_j(a^{\dagger}_j|g_j><e_j|+H.c.) \quad  0\leq j \leq N-1
\end{equation}
Here $|g_j>$ and $|e_j>$ are the ground and excited state of the $j$-th two-level atom, respectively. $\omega_j$ is the transition energy 
between the two energy levels. $g_j$ is the photon-atom coupling strength. The total Hamiltonian of the whole system is $H=H_{cv}+H_A+H_{int}$. 
Thus the stationary eigenstate may be expressed as
\begin{equation}
|\psi_k> = \sum_j\psi(j)(a^{\dagger}_j|g,0>+\psi_j|e,0>,
\end{equation}
where $|0>$ stands for the vacuum state the photon in the cavities coupled to the waveguide, and $\psi_j$, $(0\leq j \leq N-1)$, gives the 
probability amplitude of the atom in the excited state. 

\subsection{A. N atoms coupled in series}

The photon tranport in one-dimensional coupled resonator waveguide has been extensively in recent years studied with the use of scattering 
theory based on the quantum Lippman-Schwinger formalism\cite{Xu}, by transfer matrix method\cite{Tsoi}, the input-output theory \cite{Liao16} 
and the Bethe ansatz approach in the contex of the real continuous and discrete space\cite{Shen05, Zhou08}. The dynamics of the photon transport 
has be investigated by the coupled mode theory as well as the input-output theory\cite{Manolatou, XuFan}. Here we derive a formalism based on 
the tight-binding approximation\cite{Kowal}. We show that this formalism is a practical and efficient method for the calculation of the 
transmission and reflection amplitudes for arbitrary atoms and coupling modes.

Suppose that the scattering cluster is formed by $N$ cavities with identical two-level atoms inside. The TB equations for this central coupled 
lattice is described by
\begin{equation}
\omega\psi(j)-J\psi(j-1)-J\psi(j+1)+g\phi_j=E_k\psi(j),
\end{equation}
\begin{equation}
\omega_j\phi_j+g\psi(j)=E_k\psi(j),
\end{equation}
where $\omega_j$ stands for the transition energy of the coupled atoms. Taking into account the fact of identical atoms, i.e.,  
$\omega_j=\omega_0$ for $0 \leq j \leq N-1$, we find
\begin{equation}
\alpha=2\cos(k)+\frac{g^2}{\Delta_k},
\end{equation}
where $\Delta_k=\omega-2\cos(k)-\omega_0$ is the detuning. Then it follows immediately,
\begin{equation}
\begin{split}
	r = & \frac{-a_{N-1}-b_{N-1}e^{-ik}+c_{N-1}e^{ik}+d_{N-1}}{a_{N-1}+(b_{N-1}-c_{N-1})e^{ik}-d_{N-1}e^{2ik}}, \\
	t_N = & \frac{-2i(a_{N-1}d_{N-1}-b_{N-1}c_{N-1})\sin(k)}{a_{N-1}+(b_{N-1}-c_{N-1})e^{ik}-d_{N-1}e^{2ik}}. 
\end{split}
\end{equation}

For an identical coupled resonator waveguide it is shown that a finite bandwidth of perfect reflection $R=1$ may be observed for certain values of the 
atom transition frequency $\omega_0$, even for a single coupled atom. The increase of the number of coupled atom can only lead to a clear cut-off of the
band borders\cite{Chang11}. In order to demonstrate how the atom transition frequency controls the photon transport in our model system, we assume 
that the cavity-waveguide and the cavity-atom coupling strength are constant and taken to be $1$, throughout this paper. Obviously much diversified 
transport properties may emerge varying coupling constants. In general, there are two parameter regimes are considered. One is that the detuning of the 
incoming photon energy to the atom transition energy is inside the interference range, where the waveguide photon mode is comparable to that of the atom,
so that stimulated absorption or emission may occur. Another regime is the photon frequency is far away from the interference range, and so the resonant 
transmission is usually expected. In the following numerical analysis, we focus on the perfect reflection regime. This phenomenon is related to 
the manipulation of the photon propagating group velocity as a finite bandwidth makes it possible to acomodate a real wave packet. 

Fig.2 shows typical photon transport properties for identical coupled atoms. We plot the reflectance $R$ as a function of the photon wavevector and 
the detuning $\Delta$. As is shown in Fg.2 (a) and (b), there two typical parameter sets related to $R=1$. One is defined by $dR/dk=0$, and another corresponds to $dR/dk>>1$. 
Our numerical results reveals that as the atom transition energy is changed the position and the bandwidth of the $R$=1 window vary accordingly. 
As shown in Fig.2(c), when the atom frequency is changes the $R=1$ band will sweep over all spectrum and  the bandwidth increases as the atom frequency
increases from $\omega_0=0$ to $\omega_0=\omega=2\pi$. And then it decreases to its minimum at $\omega_0=4\pi$. This variation pattern repeats inself
periodically as one varies the atom frequency. It looks quite surprising that even when the incoming photon energy is complete out of the possible
interaction range with the atom, there exists still a small finite $R=1$ windows. and this leads to the atomic mirror with all-frequency perfect reflection.
It is noticed that for identical coupled atoms perfect reflection intervals can not be created by increasing the number of the atoms, which is against the 
intuitive or conclustions drawn in other model systems that the accumulated effects of the multiple scattering will result eventually to the complete
reflection[uuuuu].

\begin{figure}
\includegraphics[width=7.9cm]{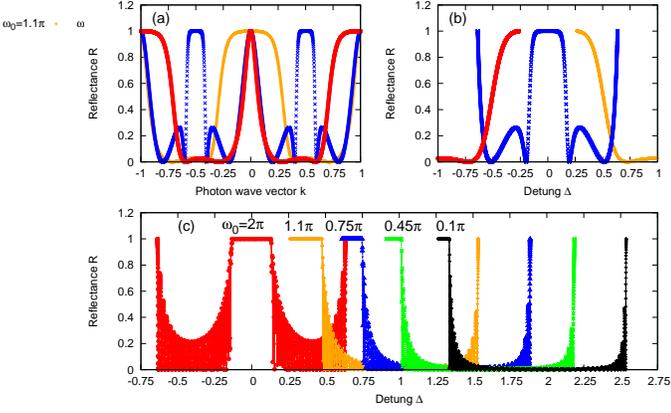}
\caption{The reflectance vs. the waveguide photon momentum of two identical atoms coupled 
in series for the atom frequency $\omega_0=2.9\pi$ (red crosses), $2\pi$(blue circles), $1.1\pi$ (orange pluses);
(b) The reflectance vs. the detuning with the same parameter values as (a).
(c) The reflectance vs. the detuning for $N=50$ atoms, with the atom transition frequency $\omega_0=2\pi$ (red circles), $1.1\pi$ (orange pluses), 
$0.75\pi$ (blue triangles), $0.45\pi$ (green crosses), $0.1\pi$ (black squres). The frequency of the waveguide photon mode is $\omega=2\pi$.}
\end{figure}

Generally speaking, there are many different coupling patterns by which the atoms are connected to form an array of
coupled resonate waveguide. In addition to the unform coupled atoms, the next important pattern
is the so-called super array of coupled resonator eaveguide, where two diferent kinds of atoms are connected alternatively, simbolized as $ABABABA...$.
Here we conduct numerical analysis on the simplest combination case, that is, two two-level atoms with different atom transition energies. Fig.3 shows
the transport features for $N=3$ atoms with connection patterns symbolized by $ABA$ and $BAB$. This type of coupling resembles the model 
systems studied in\cite{Yanik04, Zhou08a}, where it has been shown that such kind of systems support photon quasibound states and offers a mechanism 
for photon storage and control of photon group velocity. For this kind of super-array system, by simply changing the position order, one may obtain an ideal atomic
mirror with perfect reflection.

\begin{figure}
\includegraphics[width=7.9cm]{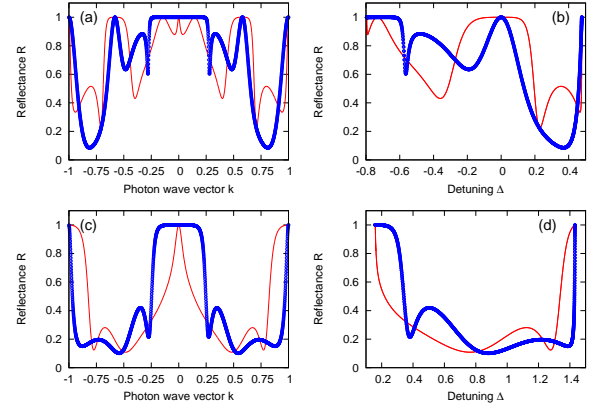}
\caption{Pattern-dependent photon transport. Here we denote by $A$ the atoms with $\omega_1$ and $B$ those with $\omega_2$. For coupled 
resonator waveguide with different atom transition frequency, we find that different sequence of serially coupled atom will lead to 
properties. Fig.3(a) and (b) shows the reflection constant $R$ of three coupled atoms with sequence $ABA$ (red line) and $BAB$ (blue circles) 
for $\omega_1=5.5$ (the atom A) and $\omega_2=3.2$ (the atom B), the frequencies of both atoms are within the interference zone with the waveguide photon, 
whose frequency is $\omega=5$. 
Fig3(c) and (d) are for $\omega_1=8.5$ and $\omega_2=2.5$, both fall outside the interaction regime with the propagating photon. }
\end{figure}

\subsection{B. N atoms coupled in parallel}

One of the well studied two-terminal parallel connected system is the quantum ring, where two paragation channels are merged at the 
left and right junctions, through which they are connected to the left and right lead, respectively. Here we first derive a formalism for an
arbitrary number of transmission lines that are joined at their extremes and coupled to two terminals (see Fig.1(b)). 

To illustrate the effects of the branched scattering processes in each single coupled waveguide, let us first study the special case 
where there is only one atom coupled to the single waveguide. Consider now $N$ resonators with embedded two-level atoms, are connected 
to the left and the right leads in a parallel way, which are labelled by $j=0$ through $j=N-1$. Those resonators are coupled to the 
same $0$-th resonator to the left and to the $N$-th oscillator to the right lead, respectively. Hence the tight-binding equations for 
those parallel connected resonators read
\begin{equation}
\omega\psi(j)-J\psi(-1)-J\psi(N)+g_j\phi_j=E_k\psi(j),
\end{equation}
\begin{equation}
\omega\psi(-1)-J\psi(-2)-J\sum_{j=0}^{N-1}\psi(j)=E_k\psi(-1),
\end{equation}
\begin{equation}
\omega\psi(N)-J\psi(N+1)-J\sum_{j=0}^{N-1}\psi(j)=E_k\psi(N),
\end{equation}
\begin{equation}
\omega_j\phi_j+g_j\psi(j)=E_k\psi(j),
\end{equation}
Using the standard traveling wave function
\begin{eqnarray}
\psi(-1) & = & e^{-ik}+re^{ik},\\
\psi(N) & = & te^{ikN}.
\end{eqnarray}
and defining that
\begin{equation}
\gamma_N=\sum_{j=0}^{N-1}\frac{J}{\omega-E_k+g_jG_j}
\end{equation}
\begin{equation}
G_j=\frac{g_j}{E_k-\omega_j}, \quad \quad 0\leq j \leq N-1
\end{equation}
we obtain the corresponding transmission and reflection amplitudes,
\begin{equation}
r=\frac{2\gamma_N\cos(k)-1}{1-2\gamma_Ne^{ik}},
\end{equation}
\begin{equation}
t_N=te^{ik(N-1)}=\frac{-2i\gamma_N\sin(k)}{1-2\gamma_Ne^{ik}},
\end{equation}

It is an easy matter to verify that if $\Delta_j=\omega - \omega_j-2\cos(k)=0$ or $\omega - E_k+g_jG_j=0$, which gives rise immediately 
to $R=1$, and $T=0$. Moreover, when $N$ goes to infinity, we have $R=\cos^2(k)$ and $T=\sin^2(k)$.

Fig.4 demonstrates the reflectance of single photon through a ring with one atom (a) and two atoms (b) on each branch. An interesting feature
of the scattering spectrum is coexistence of perfect transmission and reflection bands as illustrated in Fig.4(a). This result suggests
that coupled resonator ring waveguide behaves just like an energy filter 
that allows photons of certain energy passes freely and reflects others completely.

\begin{figure}
\includegraphics[width=7.9cm]{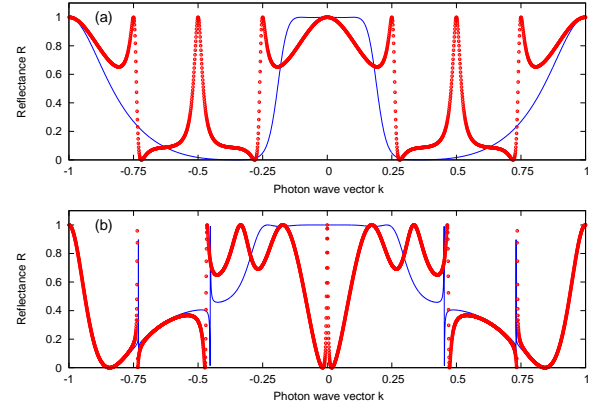}
\caption{The reflection spectra for single photon scattering by a ring, with one atom on each arm of the ring. (a) the two atoms with the 
same frequency $\omega_1=\omega_2=3.1$ (blue line), and with different frequency $\omega_1=4$ and $\omega_2=6$ (red circles); (b) the frequency of 
the two atoms are $\omega_1=4$, and $\omega_1=2$, respectively. The atom array on both branches is AB (red circles), while the array pattern is AB on 
the upper arm, and BA on the lower one.}
\end{figure}
We now turn to calculate the transport property for general parallel connected coupled resonator waveguides. In this case the scattering box is composed 
by $N$ identical parallel coupled resonator waveguides, 
where the left single waveguide is split into $N$ identical sub-waveguides. In this case we have $\psi_j(0)=\phi(0)$ and $\psi_j(n_j-1)=\phi(n-1)$. 
Assume that there are $N_0$ parallel connected waveguides, and each of them coupled with $n$ identical two-level atoms. Thus, (5) becomes
\begin{equation}
\begin{split}
	1+r = & \sum_{i=1}^{N_0}\psi_i(0)=N_0\phi(0), \\
	t_N = & \sum_{i=1}^{N_0}\psi_i(n_i-1)=N_0\phi(n-1). 
\end{split}
\end{equation}
By employing the tight-binding equations, we obtain the following results,
\begin{equation}
\begin{split}
	r = & \frac{-a_{n-1}-N_0b_{n-1}e^{-ik}+N_0c_{n-1}e^{ik}+N^2_0d_{n-1}}{a_{n-1}+N_0e^{ik}(b_{n-1}-c_{n-1}-N_0d_{n-1}e^{ik})}, \\
	t_N = & \frac{-2iN_0(a_{n-1}d_{n-1}-b_{n-1}c_{n-1})\sin(k)}{a_{n-1}+N_0e^{ik}(b_{n-1}-c_{n-1}-N_0d_{n-1}e^{ik})}. 
\end{split}
\end{equation}
For $N_0=1$ we return to the case of a single waveguide with serially coupled atoms. There are several remarkable features of our model system. 
One is the aparently surprising conclusion that when $N_0$ is sufficiently large, the parallel coupled system behaves like a total perfect 
reflection mirror with $R=|r|^2=1$ and $T=|t|^2=0$, independent of the the detuning. Particularly when the detuning is such that no aparent 
radiation interaction between the photon and the atom, in which case the resonant transmission is expected. The possible implication is directly 
related to desctructive interference among the photon states propagating along different channels. It is obvious that the mechanism behind the 
perfect reflection revealed here seems to be different from the serially coupled waveguide as illustrated in the literature\cite{Xu, Chang11}, 
where the total reflection is accumulation process due to multiple consecutive scatterings. Nevertheless, as will be shown later, this phenomena 
indeed is closely related to the characteristic features of the single coupled waveguide.

In Fig.5 we show that the reflectance for single coupled resonator waveguide (a) and (b), and the parallel connected multiple coupled resonator waveguides.
we find that by increasing the number of connected waveguides the perfect reflection windows expand into all spectrum, while the the bandwidth and spectrum position
are almost independent of the number of serial coupled atoms along the waveguide. 

This seems quite contra-intuitive, since the resistence in serial circuit is espected to grow and reduces in parallel connections.

\begin{figure}
\includegraphics[width=7.9cm]{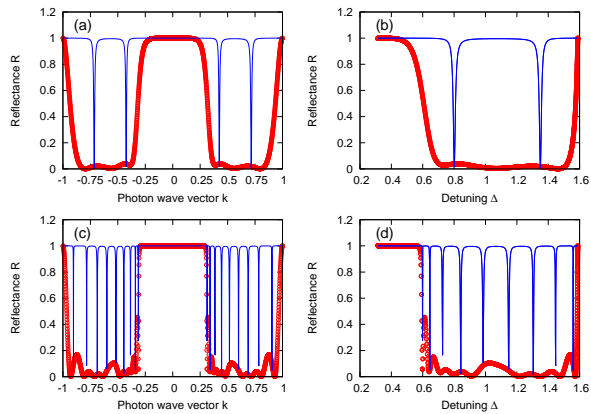}
\caption{The scattering cluster is formed by $N_p=30$ parallel connected one-dimensional cpupled resonator waveguides. The photon frequency 
is $\omega=5$, and the atom energies are $\omega_1=2$ and $\omega_2=3$. The reflection constante as a function photon wavevector (a) and 
detuning (b) for the single waveguide coupled with 3 atoms, while *c) and (d) are for 11 coupled atoms in a single channel. One of the most 
notable features is that the parallel connection tends to block out the propagation of the single photon for almost all the incoming phton 
frequency, while the increase of coupled atoms in a single waveguide can generate a fine borders of the already existing perfect reflection 
window. Note that here both atoms are outside the interference interval.}
\end{figure}

It should be emphasized that the mechanism of atomic mirror observed here is different than those discussed in Ref.\cite{Xu}, 
where the perfect reflections is attributed to the multiple collisions with serially coupled atom-contained cavities. Numerical calculations 
reveal that in contrast to the serially coupled resonator waveguide, by incorporating more cavities into the system destroy the above-mentioned 
perfect transmission and reflections windows, leaving the usual transmission patterns. This can be derived directly from Eq.(28). That is, for 
identical cavities, when $N \rightarrow \infty$, $R=\cos^2(k)$.

\section{VI. Conclusions}

In conclusion, we develop a simple tight-binding formalism for both serially and parallel connected coupled resonator waveguides, and conduct 
a series of numerical calculation of single phton transport properties. Our numerical results reveal the following novel features of the coupled resonator
waveguide: (i) For single waveguide with serially coupled cavities with embedded tow-level quantum system, the waveguide photon mode with arbitrary 
wavevector can be completely reeflected by properly chosen atom transition energy. A smilar atomic control of photon transfer can be achieved by the 
so-called super-array wavegudide where two types of atoms are connected alternatively. (ii) In the case of parallel coupled resonator waveguides, an all-frequency
ferfection reflection regime is reached when the number of coupled waveguides are sufficiently large. A photon filter can be obtained when the single photon propagate 
through a ring of coupled resonator waveguide, in which case the resonant transmission occurs for certain values of photon energys, while perfect reflection
is imposed on single photons with different wavevectors.

It is obvious that our method provide a straightforward means for investigationg one-dimentional photon transport through a scattering clusters with
certain regular internal structures, both in theoretic discussion and numerical analysis. It is expected that more real, complicated parameter regimes 
can be studied with  the use of the tight-binding formalism presented in this work, and much novel transport properties can be revealed.



\begin{thebibliography}{99}
\bibitem{Srini}

K. Srinivasan and O. Painter, Nature (London) 450, 862 (2007)
\bibitem{Raimond}

J. M. Raimond, M. Brune, and S. Haroche, Rev. Mod. Phys. 73,
565 (2001).

\bibitem{Pelton}
M. Pelton, C. Santori, J. Vuckovi´c, B. Zhang, G. S. Solomon, J.
Plant, and Y. Yamamoto, Phys. Rev. Lett. 89, 233602 (2002).

\bibitem{Bermel}
P. Bermel, A. Rodriguez, S. G. Johnson, J. D. Joannopoulos,
and Marin Soljaci´c, Phys. Rev. A 74, 043818 (2006).

\bibitem{Menon}

V. M. Menon, W. Tong, F. Xia, C. Li and S. R. Forrest,Opt.  Lett. 29, 513 (2004)

\bibitem{Astaf}

O. Astafiev, A. M. Zagoskin, A. A. Abdumalikov Jr., Yu. A.
Pashkin, T. Yamamoto, K. Inomata, Y. Nakamura, and J. S.
Tsai, Science 327, 840 (2010).

\bibitem{Shen05}

J.-T. Shen and S. Fan, Optics Lett. 30, 2001 (2005).

\bibitem{Shen07}
J.-T. Shen and S. Fan, Phys. Rev. Lett. 98, 153003 (2007).

\bibitem{Gardiner}

C. W. Gardiner and M. J. Collett, Phys. Rev. A 31, 3761 (1985).

\bibitem{Fan10}

S. Fan, S. E. Kocabas, and J. T. Shen, Phys. Rev. A 82, 063821 (2010).

\bibitem{Fan99}
S. Fan, P.R. Villenueve, J.D. Joannopoulos, M.J. Khan, C. Manolatou, and M.A. Haus,S. E. Kocabas, and J. T. Shen, Phys. Rev. B 59, 15 882 (1999).

\bibitem{Xu}
Y. Xu, Y. Li, R. K. Lee, and A. Yariv,
Phys. Rev. E 62, 7389 (2000).

\bibitem{Tsoi}
T. S. Tsoi and C. K. Law, Phys. Rev. A 80, 033823 (2009).

\bibitem{Roy}
D. Roy, Phys. Rev. Lett. 106, 053601 (2011).

\bibitem{Zhou08}
L. Zhou, Z. R. Gong, Yu-xi Liu, C. P. Sun, and F. Nori, Phys.
Rev. Lett. 101, 100501 (2008).

\bibitem{Yanik04}

M.F. Yanik, W. Suh, Z. Wang, and S. Fan, Phys. Rev. Lett. 93, 233903 (2004).

\bibitem{Yanik05}

M.F. Yanik and S. Fan, Phys. Rev. A 71, 013803 (2005).

\bibitem{Zhou08a}

L. Zhou, H. Dong, Yu-xi Liu, C. P. Sun, and F. Nori, Phys. Rev. A 78, 063827 (2008).

\bibitem{Chang11}

Y. Chang, Z.R. Gong, and C.P. Sun, Phys. Rev. A 83, 013823 (2011).

\bibitem{Zhou13}

L. Zhou, L. Yang, Y. Li, and C.P. Sun, Phys. Rev. Lett. 111, 103604 (2013).

\bibitem{Cheng12}

M. Cheng, X. Ma, M. Ding, Y. Luo, and G. Zhao, Phys. Rev. A 85, 053840 (2012).

\bibitem{Liao16}

Z. Liao, H. Nha, and M. S. Zubairy, Phys. Rev. A 93, 033851 (2016).

\bibitem{Qin16}

W. Qin and F. Nori, Phys. Rev. A 93, 032337 (2016).

\bibitem{MZI}

Yi Xu and Andrey E. Miroshnichenko,
Phys. Rev. A \textbf{84}, 033828 (2011).

\bibitem{Calajo}

G. Calajo and P. Rabl, 	arXiv:1612.06728 (2016).

\bibitem{Kowal}

D. Kowal, U. Sivan, O. Entin-Wohlman, and Y. Imry, Phys. Rev. B 42, 9009 (1990).

\bibitem{Manolatou}

C. Manolatou, M.J. Khan, S. Fan, P.R. Villenueve, and M.A. Haus, IEEE J. Quan. Elec. 35, 1322 (1999).

\bibitem{XuFan}

S. Xu and S. Fan, Phys. Rev. A 94, 043826 (2016).

\end{thebibliography}
\end{document}